\documentclass[Afour,Sagev,times]{sagej}

\usepackage{moreverb,url}
\usepackage[utf8]{inputenc}
\usepackage[english]{babel}
\usepackage{soul}

\usepackage[dvipsnames]{xcolor}
\usepackage{color}
\definecolor{darkgreen}{rgb}{0,0.5,0}
\definecolor{purple}{rgb}{1,0,1}
\definecolor{orange}{rgb}{1.0, 0.5, 0.0}
\newcommand{\kibitz}[2]{\ifnum\Comments=1\textcolor{#1}{#2}\fi}

\usepackage{xcolor}
\usepackage{fontawesome5}
\usepackage{float}

\definecolor{orcidlogocol}{rgb}{0.65, 0.807, 0.223}
\newcommand{\orcid}[1]{$\,$\href{https://orcid.org/#1}{\textcolor{orcidlogocol}{\faOrcid}}}

\newcount\Comments  
\usepackage[colorlinks,bookmarksopen,bookmarksnumbered,citecolor=red,urlcolor=red]{hyperref}

\newcommand\BibTeX{{\rmfamily B\kern-.05em \textsc{i\kern-.025em b}\kern-.08em
T\kern-.1667em\lower.7ex\hbox{E}\kern-.125emX}}

\def\volumeyear{2021}

\begin{document}

\runninghead{Andrea Russo et al.}

\title{Entropy-rate as prediction method for newspapers and information diffusion}

\author{Andrea Russo\affilnum{1-2}, Antonio Picone\affilnum{1}, Vincenzo Miracula\affilnum{1}, Giovanni Giuffrida\affilnum{3},  Francesco Mazzeo Rinaldi\affilnum{3-4} }

\affiliation{\affilnum{1} Department of Physics and Astronomy, University of Catania, Italy\\
\affilnum{2} Department of Sociology "Newman", University College Dublin, Ireland\\
\affilnum{3} Department of Political and Social Science, University of Catania, Italy\\
\affilnum{4} KTH, Royal Institute of Technology, School of Architecture and the Built Environment, Stockholm.}

\corrauth{Andrea Russo,
Department of Physics and Astronomy "Ettore Majorana", 
University of Catania,
Via S. Sofia, 64
95123 Catania (Italy)}
\email{Andrea.russo@phd.unict.it}

\begin{abstract}
This paper aims to show how some popular topics on social networks can be used to predict online newspaper views, related to the topics. Newspapers site and many social networks, become a good source of data to analyse and explain complex phenomena. \\
Understanding the entropy of a topic, could help all organizations that need to share information like government, institution, newspaper or company, to expect an higher activity over their channels, and in some cases predict what the receiver expect from the senders or what is wrong about the communication.\\
For some organization such political party, leaders, company and many others, the reputation and the communication are (for most of them) the key part of a more and complex huge system. \\
To reach our goal, we use gathering tools and information theory to detect and analyse trends topic on social networks, with the purpose of proved a method that helps organization, newspapers to predict how many articles or communication they will have to do on a topic, and how much flow of views they will have in a given period, starting with the entropy-article ratio. \\
Our work address the issue to explore in which entropy-rate, and through which dynamics, a suitable information diffusion performance is expected on social network and then on newspaper. 
We have identified some cross-cutting dynamics that, associated with the contexts, might explain how people discuss about a topic, can move on to argue and informs on newspapers sites. 

\end{abstract}

\maketitle

\section{Introduction}


Nowadays public discourse around public and political events, especially those that occur online, are increasingly influenced by the issues that media companies choose to concentrate on \cite{R1}. News media not only have the power to create public awareness around social issues \cite{R2} \cite{R3}, but they can also influence how the public perceives what issues are most important \cite{R4} \cite{R5}.  These dynamics are essential to understanding how people consume news and how they engage in discussions around specific events. 
Social media are the most relevant online communication channels designed for networking \cite{R6}. When used effectively, their applications can promote dialogue \cite{R7}, facilitate information transfer and understanding \cite{R8}, engage stakeholders  \cite{R9}, and improve communication and collaboration in online environments \cite{R10}. \\ 
Spreading the right information to the right audience has always been the underlying goal of proper communication. However, it is hard to estimate whether the intended addresses of the communication have understood the information correctly, especially if the process of communication was irregular and complex and/or the event to communicate is complex. 
Today, we believe, it might be possible to assess whether the addressees have received the sender's right information by analyzing the related data created on the various digital platforms, and advising the newspaper itself about the possible amount of visualization from that specific topic on his site.
To make it clear, when the addresses have received the right information, there is no need for other information to satisfy the addressees about the event \cite{R34}. 


From a methodological standpoint, users on social networks often use \textit{hashtags} embedded in their discussion. Thus, we use hashtag-related data to estimate any lack of understanding or curiosity about a particular issue: if people continue to talk about a specific topic, they unconsciously need better explanation and information on that topic \cite{R11}. This can be measured by analyzing readers’ activity within a specific time frame. 

We believe that is possible to properly measure real-time hashtag trends on social networks and information entropy-rate —for specific topics—  by informing the addressees about the inferred and perceived information by the readers.

In general, this is very handy for political communication or business news when prompt public reactions to given issues are particularly relevant. The entropy-rate is the hypothetical average level of "information", "surprise", or "uncertainty" inherent in the variable's possible outcomes, and we use it to estimate complex processes like the reception of information. 
We use data from social networks to study the crucial factors in the communications process. We measured the natural attention/interest temporal decay for some topics and used these curves as a benchmark in our study. In general, we noticed that standard news, such as sports events or flash news, exhibits an interest natural temporal decay measured by a low entropy value and standard behavior. Conversely, high entropy values for some topics lead to an \textit{unbalance} between intended information (by publishers) and perceived information (by readers). 

We applied our model to different, and sometimes complicated topics, such as government/institutional communication, Political news, Global news, Disaster news, Sport news and Daily news. 
In general, some of those news have been followed by a low social network activity, but other have shows an intense social network activity, due to the high entropy about the topic.   

This paper aims to contribute to such discussion by using social networks data through an entropy-based model and computational social science approach.
Our model does not intend to understand the entropic reaction to different communication channels, but our model is to examine the entropic reaction to the type of event (standard and equal to all other communication channels), and predict circa the amount of online traffic on online newspapers.   \\
\\
This paper is organized as follows: Section 2 introduce relative works on information entropy, Section 3 describes the methodological approach, while Section 4 presents the case study, Section 5 show the data, result and the information structure, and finally section 7 conclusions and future work. 

\section{Related works}
Entropy-based analysis and computational method have already been introduced in various prediction scenarios such as atmosphere \cite{R30}, network traffic \cite{R31}, human mobility \cite{R32} and Radio Spectrum State dynamics \cite{R33}. 
\\
Many of these related works focus on public opinion research, influence cascade, factors influencing diffusion of emergency information, and many others. 
Our paper tries to advance the research on the process of information diffusion, adding newspaper data to evaluate the uncertainty of the information and the request from people to further/extra information.\\
We try also, to improve the research on emergency information dissemination \cite{R21} with an entropy classification, enlighten that in the field of social media research many other topics and events can be studied and kept under observation, and in some cases be predictable. 
\\


Siqing Shan and Xiao Lin \cite{R21}, have established how to efficiently evaluate uncertainty, with the Ya’an earthquake and Wenchuan earthquake, revealing information’s uncertainty in population.
The analysis of relevant blog content revealed that when the Ya’an earthquake was mentioned, it was often associated with the Wenchuan earthquake, which occurred five years earlier. Some people even considered the Ya’an earthquake an aftershock of the Wenchuan earthquake, thus prompting a revival of the topic of the Wenchuan earthquake. 



Yin Jie et al. \cite{R22}, have analyzed Twitter messages generated during humanitarian crises, and focused on the classification of the disaster event (e.g., traffic accidents and civil disorders). 
The result showed that more generic features, such as hashtag count or mention count, were more effective than incident-specific features, such as actual hashtag or mention values, for identifying previously unseen types of disasters. Accuracies for the best setting (unigram plus hashtag count) varied between 60\% to 73\%. Also they monitored Twitter for a specific event, evidencing that a large volume of tweets published every second are considered irrelevant. 
Even when the tweets are discussing an event of interest, depending on the application one may be interested in prioritizing different classes of messages. To address this need in the context of disaster management, they study three different tweet classification settings: disaster or not, disaster type, and impact assessment.  

Peng Sancheng et al. \cite{R23} presented a framework to quantify social influence in mobile social networks. The social influence of users was measured by analyzing the SMS/MMS-based communication behaviors among individuals.
In addition, they revealed and characterized the social relations among mobile users through the analysis of the entropy of friend nodes and the entropy of interaction frequency. 
The analytical results show that the influence spread of their proposed method outperforms that of the random method and that of the degree-based method. Also, they show how only the top 1 to 5 nodes are influential nodes and the succeeding nodes do not contribute to increasing the influence spread. 
  

Kolli et al. \cite{R24} have studied and quantified the interplay between the regular (and thus predictable) and the random (and thus unforeseeable) underlying online social media cascade dynamics.
Their research aims to apply a framework for characterizing the predictability of cascade trajectories and the theoretical “maximal predictability". 
They show that at least 20\% of the time the cascade volume changes in a manner that appears to be random, and in the remaining 80\% of the time it is possible to predict the cascade’s future volume. 

Borge-Holthoefer et al. \cite{R25}  have used entropy analysis on information-driven dynamics. 
They show up that in some cases, discussions evolve organically, building up momentum up to the point where the exchange of information is generalized; however, in some other cases, the discussions emerge suddenly as a reaction to some unexpected external event. They work focusing on protests or breaking-political events (Brazilian protests) or public debate. 



Barros et al. \cite{R27} propose a novel method to detect events on Twitter based on the calculation of the entropy of the content of tweets in order to classify the most shared topic as an event or not. 
Their goal consists of capturing the phase transition of entropy when it changes to a certain value. They hypothesized that during the occurrence of an event, the entropy of the bigrams extracted from the social media changes its dynamics, they observed a continuous phase transition of the entropy dynamics. 
Furthermore, this provides some evidence that their method is sensitive to detecting events that occur in a short time. He obtained data from a football match, where breaking-event (penalty, yellow or red card) can appear suddenly provoking reactions on social networks.

Park Han Woo  \cite{R28}, has proposed negative entropy not as a comprehensive or representative index of elections but as an experimental and innovative measure for events occurring in social media environments.
The research question was 1) What (social) media are more likely to generate (negative) entropy across different periods? ; 2)  Which politicians (or pairs of politicians) are more likely to generate (negative) entropy for bilateral, trilateral, or quadruple relationships across various media and periods?
He examines the information entropy produced by various types of web-mediated and social media platforms before the election, particularly with respect to presidential candidates. Twitter showed the greatest negative entropy, followed by Facebook and Google, in that order.




Senevirathna et al \cite{R29}. present a novel method for tracking influence relationships over time and linked influence cascades. 
Focusing on cryptocurrency communities, they show that user distributions over influence cascades for the cryptocurrency community were robust across platforms, while others community were more platform-sensitive.
They evidence that influence cascades are typically analyzed using a single metric approach (all influence is measured using one number). However, social influence is not monolithic; different users exercise different influences in different ways, and influence is correlated with the user and content-specific attributes. 
Also, they evidence how influence was propagating within the cryptocurrency community. On both Twitter and GitHub they were similar, with the exception of contribution and sharing events. Furthermore, there are no similarities in how influence is propagated when comparing the two communities on either platform (Twitter and Github).
In sum, they illustrate how to entropy can be used as the measurement of influence to estimate the degree to which causal relationships existed between user actions, and they do not find any individual influence relationships across the two communities on Twitter that show significantly similar progressions of the magnitude of influence over cascade level. \cite{R29}

\section{Methodological approach}

To evaluate and predict information dynamics on social networks and newspapers, we used several tools to justify the data quality and the dynamics on different platforms. In detail, we have described the tools and their goal in Table \ref{tab:1}.

\begin{table}
    \small\sf\centering
    \begin{tabular}{c|l}
    \hline
    Tools & Goals \\
    \hline
    Twitter API & Collect data from Twitter \\
     Shannon Entropy   &  Evaluate High or Low entropic events \\
     Info structure & Analyse data sample\\
     Scraper  &    Collect data from Newspaper \\
     Data-viz  & Visualize data-patterns \\
     Odds-ratio & Correlation between topic and articles\\
     \hline
    \end{tabular}
    \caption{Tools \& Goals}
    \label{tab:1}
\end{table}

\subsection{Twitter API}
The Twitter API enables programmers and researchers access to Twitter elements like Tweets, Direct Messages, Spaces, Lists, users, and more.
We have collect tweets related to various topics, by using Tweepy and the Twitter archive API.
Both services use permission from Twitter to obtain and gather data, but any downloaded topic needs revisions and a cleaning process to increase the quality of the research. For example, we found many copy-paste tweets (caused by spamming process, or fake-account/bot), and also several tweets had repeated questions marks, they were thus removed. 
For any topic, we use the same methodology to obtain standard and quality data. This is request also from the entropy calculation because it is essential to know every evolution of the topic in every hour. 
In addition, to obtain the correct amount of tweets (defined as the number of tweets) for each day/hour we use getdaytrends.com, a specific site where it is possible to monitor every topic in real-time and also aged topic.
Also, the Topic selection process, as follows various information categories from newspapers. We select specifically: Political topics, World News, Sports News, Cultural news, and City news. 
In total, our data count more than 20.000 tweets.
\subsection{Shannon Entropy}     
We used Shannon entropy, to evaluate and quantify the degree of information assimilation from people about topics. 
In information theory, the “importance of the information” contained in a message is directly related to how “hits” or “sudden” the message is for the reader \cite{R12} \cite{R13}.
Suppose we have a biased coin with probability $p$ of landing on heads and probability $1 - p$ of landing on tails. For what value of $p$ do we have the maximum "surprise" or "uncertainty" on the outcome of a coin toss? If $p = 1$, the outcome of a coin toss is expected to get always head, so there is no surprise or uncertainty. Similarly for $p = 0$, when we always expect the coin to land on tails. If $p = 0.5$, then we have the maximum surprise or uncertainty.
This value of hits or sudden has been mathematically proposed by Claude Shannon \cite{R13} in 1948 as part of his theory of communication. After a discussion with John Von Neumann, Shannon decided to use the term "entropy" in place of the word "uncertainty". At a conceptual level, Shannon's Entropy is the "quantity of information" of a variable.
This turns into the amount of memory (e.g. number of bits) required to store the variable, which can be understood as the amount of information contained in that variable. Shannon's entropy is expressed by a number and the calculation is straightforward; in particular, it is not the number of bits needed to represent all the different values that a variable could take, but it is just the raw data.

Shannon's intuition was that "less storage might be sufficient to store the information".
Shannon's entropy metric helps to identify the amount of storage space needed for information. An alternative way of looking at entropy is therefore as a measure of the "compressibility" of the data, that is, a compression metric. However, how far can the raw data be compressed without losing the information?

Formally, the entropy of our biased coin is given by:
\begin{equation}
   H(coin) = -(p * log(p) + (1-p) * log(1 -p))
\end{equation}
and the base of the logarithm can be chosen arbitrarily. When the base is 2, the entropy is measured in bits. If instead the base is \textit{e}, the entropy is measured in \textit{nats}. Finally, if the base is 10, the entropy is measured in \textit{dits}.

When $p = 0.5$, $H(coin)$ is maximal, and it is equal to 1 bit. When instead $p = 0$ or $p = 1$, $H(coin)$ is minimal, and it is equal to 0 bits.

The concept of entropy can be generalized from the simplest case of a coin to the more complex case of a discrete probability distribution. A discrete probability distribution over $n$ possible outcomes $x_1, ..., x_n$ is given by $n$ probability values $p(x_1), ..., p(x_n)$, 
where $0 <= p(x_i) <= 1$, and the sum of all $p(x_i)$ is equal to 1. Note that a coin is a probability distribution over two possible outcomes.

Formally, the entropy of a discrete probability distribution is defined by:
\begin{equation}
  H(P) = - (p_1 * log(p_1) + ... + p_n * log(p_n))
\end{equation}
where, as before, the base of the logarithm can be chosen arbitrarily.

A clearly communicated concept does not lead to a “surprised” or “unexpected” reaction: its information value is low, which corresponds to a low entropy level. Conversely, a poorly communicated concept leads to high entropy (i.e., a more surprising effect) on the readers. A poorly understood message triggers a sudden need in the reader for additional information for clarification and better understanding, for example, students asking for more information and clarification to understand complex topics. 

\begin{figure}
	\centering
		\includegraphics[width=\linewidth]{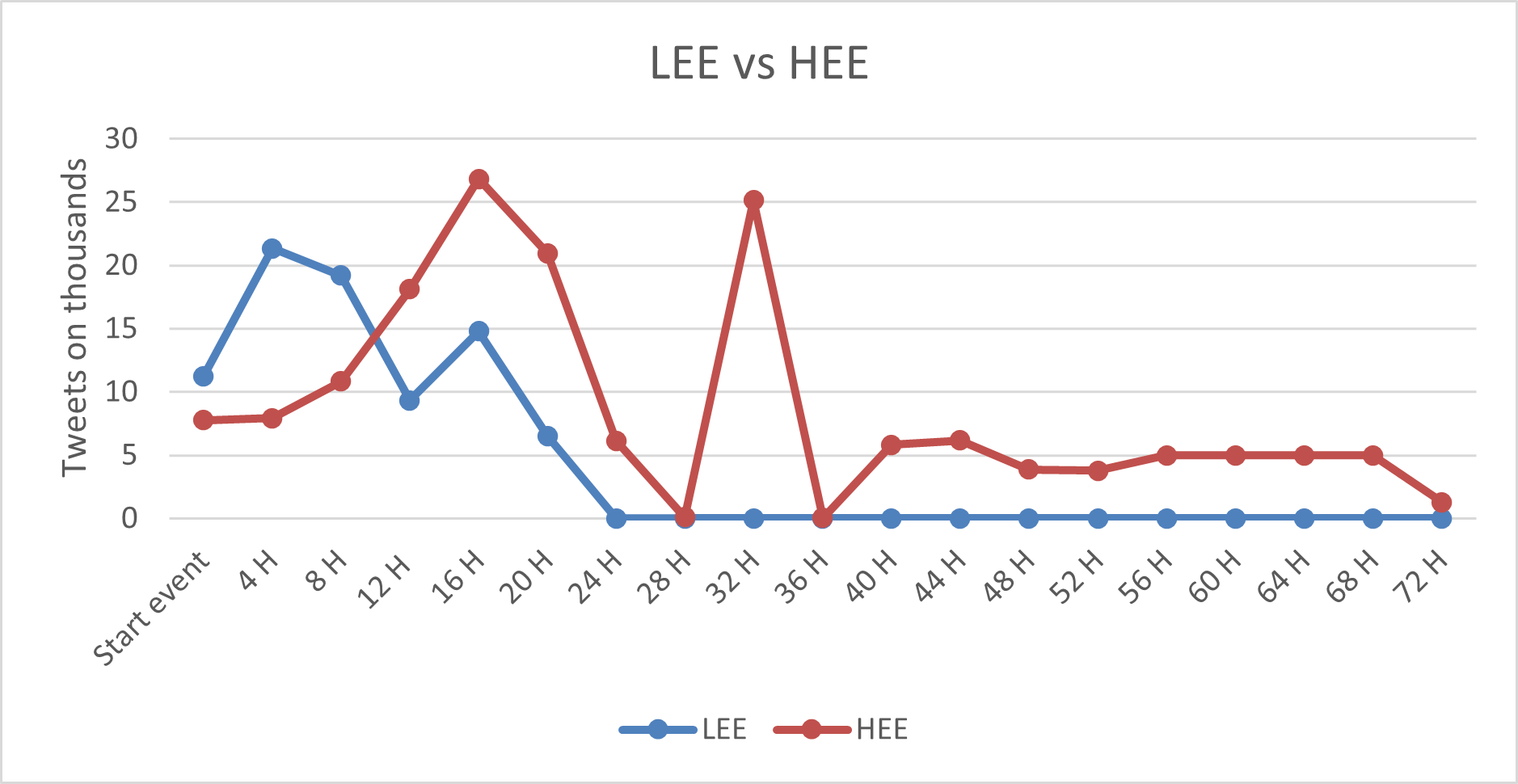}
	\caption{Entropy events (LEE) vs High entropy events (HEE) median for 72 hours and relative tweets}
	\label{FIG:1}
\end{figure}

In today’s digitized world, this information request produces a hype of digital activity such as social network discussion, Google searches, newspaper consumption, etc. Thus, by real-time measuring such digital activity, we might estimate the entropy of a specific topic and, this allows us to recognize its efficiency. 
Assessing the information diffusion quality of a certain topic with the entropy-rate depends on the information entropy level of the message which, we show, is directly related to the addressees’ reaction to the message. We measure addressees’ responses by analyzing social media data (in the same period) connected to that topic.\\
We believe that if a topic is well-communicated, people do not (necessarily) need to comment about it on Twitter or, at least, no more than usual. Otherwise, as mentioned above, if the message was unclear, people will look for better understanding by twitting comments and expressing opinions on the Web. The more social media activity on that topic we measure, the higher entropy for that topic is. \\ 
For the purposes of our analysis, we distinguish two types of events: 
\begin{enumerate}
\item Low entropy events (LEE) 
\item High entropy events (HEE)
\end{enumerate}
LEE are events whose, the information diffusion was quickly understood by the public.  
In general, immediately after news on this type of event appear either in newspapers or on TV, there is a hype on social media activity. LEE social media-related activity usually decays almost entirely in 4 hours. This holds true in general for many easy-to-understand topics like sports events, flash news, natural routines, etc. Information on such types of events shows low entropy levels and standard behavior. 
Conversely, HEE are more complicated events. Information dissemination of these events is inherently more difficult compared to LEE. Social media activity for this type of communication follows an entirely different pattern. We measure a decay curve over a 24-72 hours period. During this time, an intense communication activity takes place on social media. HEE exhibit a higher entropy level in general as their communication complexity is higher. After that time, people tend to stop discussing and commenting on those.  
In Figure \ref{FIG:1} we show the different trends for LEE and HEE (Thousands of tweets). 

We have identified and analyzed many different topics-spawn events on social networks like everyday events (sports, weather, etc.), cyclic-time events (Mondays-motivation, etc.) as LEE, and unpredictable events like uncommon events (an important unexpected murder; e.g., Jeffrey Epstein), and uncommonness outcomes events as HEE (see fig. 1). High entropy information produces a better information diffusion than Low entropy one; but also, people would have difficulties understanding it. In contrast, Low entropy information requires less effort to be understood and hard effort to be spread. These events' natural decay shows a complicated and long curve that grows and slows down during day-nighttime, but it stays a trending topic for at least 60 hours. Other natural decays show fewer dynamics and more predictive behavior. HEE remains rare, but they indicate the unbalance between the senders' information requested and the information supply. How is it possible to estimate if something is wrong, and how is it possible to expect high entropic events?   

The time when everyone listens and tries to assimilate the information is a crucial stage because if the communication goes well (sound information diffusion and right information received), there is no HEE. Still, if the message/information's entropy is high, the HEE arises (1). After that, usually, people start to debate and talk on social networks and (or) in our private circle (family, friends, etc.). 
At this point, if enough people start to speak and the topic became a Trend topic for a long time (at least 40 hours), we can evidence how much the information diffusion was enlarged and how much the information was clear (2). At some point, the organization that has provoked the HEE will dispatch the correct definition of the information sends (3).
\subsection{Information structure}
The Information Entropy of a message can be improved with an information structure method. To prove if the HEE are more entropic and consequentially need more information to be assimilated, we have chosen to analyze the syntax structure sample from Twitter.  
Other papers evidence a significant change related not only to the entropy rate, but also to the information structure on social networks.\\
The Arabic Newspapers work \cite{R18}, had tried to recognize significant changes in the social state via changes in the linguistic performers of the media. Entropy rate, and the information structure about Hashtags on social networks can also describe cultural and socials dynamics, also adding information request coming from society. 
This data also suggests a strategic way to improve the diffusion of the information; the entropy rate of some information (especially institutional and political information and/or in emergency case \cite{R20}) 
should not be at the medium between HEE and LEE, but tactically it should be slightly above the average, so as to be compressible to many people, and at the same time, alluring, because the average information spread is more entropic. 
Also, Minkyoung et al. (2013) \cite{R19}, have studied the dynamics of information diffusion across social networks, confirming this logic, showing topics with controversial subjects, which seem to lead to longer discussions representing personal opinions, like the HEEs case. 
\subsection{Scraper}
We use a homemade scraper to collect data from newspapers like articles, name of the article, the text of the article, and numbers of visualizations by hours/days. 
Those data give us the resources to quantify the interaction and the dynamic between readers and the most relevant topic of the day.\\
As for the API methodology, we have collected, cleaned, and supervised the data to increase the quality of the data and research. 
The revisions and the cleaning process has taken some time, due to the unstructured data gathering. 
\subsection{Data-viz}
Data-viz isn't a single tool, but a collection of tools to elaborate data and visualize them.
With data visualization, it is possible to translate information into a visual context, such as a map or graph, to make data easier for the human brain to understand and pull insights from. 

We have used many different simple or complex data visualization tools, but in the end, we used the simple and original template of the data visualization tool available in Excel. \\
Thanks to the data collected from Twitter API and the Scraper, we merge the data to observe oscillation and information dynamics between Social networks and newspaper digital platforms.
\subsection{Odds-ratio}
The odds ratio is a statistical measure that shows the degree of correlation between two factors. Odds ratios are used to compare the relative probabilities of the event of interest, given the exposure to the variable of interest. 

The odds ratio helps to identify the probability that an event 'A' leads to a specific event 'B'. The higher the odds ratio, the greater the probability of the event occurring with the exposure. Odds ratios of less than one imply that the event is less likely to be related. 

\section{Data \& Case study}
During 2020, given the high amount of atypical events, we have collected many Tweets related to various topics. 
We have used Shannon's Entropy to evaluate the request for information from people, Table \ref{tab:2} shows the HEE and LEE between January 8th 2020 to November 1st 2020. Meanwhile, Figure \ref{FIG:1} shows the average time dynamic activity between HEE and LEE.
\footnote{"Times by hours" define how many hours the topic is "alive" since his first apparition on social network and newspaper; \\ "Activity on internet" define how many tweets or visualization are related to the specific topic. 
Obviously, the total amount of tweets is bigger then the newspaper visualization, so we had to reduce by 10, to simplify and ease the socialnetwork-newspaper dynamic. }\\
By real-time measuring information entropy on social media, we could promptly get informed how easy to understand a certain communication was. And we believe this may have various useful applications in real life.

\begin{table}
\small\sf\centering
\caption{Information Entropy and average tweets/hours.}
\begin{tabular}{|l|l|c|c|c|}
\hline
Type & Topic &Tweets &Hours& Entropy\\
\hline
HEE &	 \texttt{DPCM (Apr 26th)}&	894.500&	78&	\textbf{5.6113}\\
HEE & \texttt{BLM}&8.4 M & 127 & \textbf{5,5760}\\
HEE & \texttt{Beirut}&3.2 M & 69 & \textbf{5,1962}\\
HEE & \texttt{Debates2020} &3.7 M & 47 & \textbf{4,8745}\\
HEE & \texttt{Megxit} & 52.100 & 63 & \textbf{4,3127} \\
HEE &	 \texttt{DPCM (Oct 25th)} &	618.150	&42	& \textbf{4.1972}  \\
& & & & \\
LEE & \texttt{Int.WomensDay} & 1.5M &21 & \textbf{4,1457} \\
LEE & \texttt{ValentinesDay} & 812.900 & 34 & \textbf{4,0572} \\
LEE & \texttt{Tokyo2020} & 3.1 M & 24 & \textbf{3,6664} \\
LEE &	 \texttt{SuperLega} &	45.400&	27&	\textbf{4.0541}\\	
LEE &	 \texttt{G.dellibro} &	11.600&	14&	\textbf{3.8232}\\	
LEE &	 \texttt{JuveInter} &	48.500&	16&	 \textbf{3.6126} \\	
LEE &	 \texttt{MilanNapoli} &	16.800&	10&	\textbf{3.5672}\\	
LEE &	 \texttt{DPCM (Oct 13th)} &	26.700&	12	&\textbf{3.1569}\\	
LEE &	 \texttt{Dupasquier} &	 <10.000&	7&	 \textbf{3.0201}\\	
LEE &	 \texttt{GazaU.Attack} &	623.800	&8	&   \textbf{2.9740}\\	
LEE &	 \texttt{ObiWan} &	20.300	&7	 & \textbf{2,6907}\\
LEE &	 \texttt{SuperLeague} &	616.700	&48	 & \textbf{2.5417}\\
\hline
\end{tabular}
    \label{tab:2}
\end{table}

We have chosen to study the HEE because they are not only demonstrating higher activity in social networks but also because there is significantly more discussion (debates) on HEE topics than LEEs. In this chapter, we would make a brief description of the HEEs chosen for our work.

\begin{figure}
	\centering
		\includegraphics[width=\linewidth]{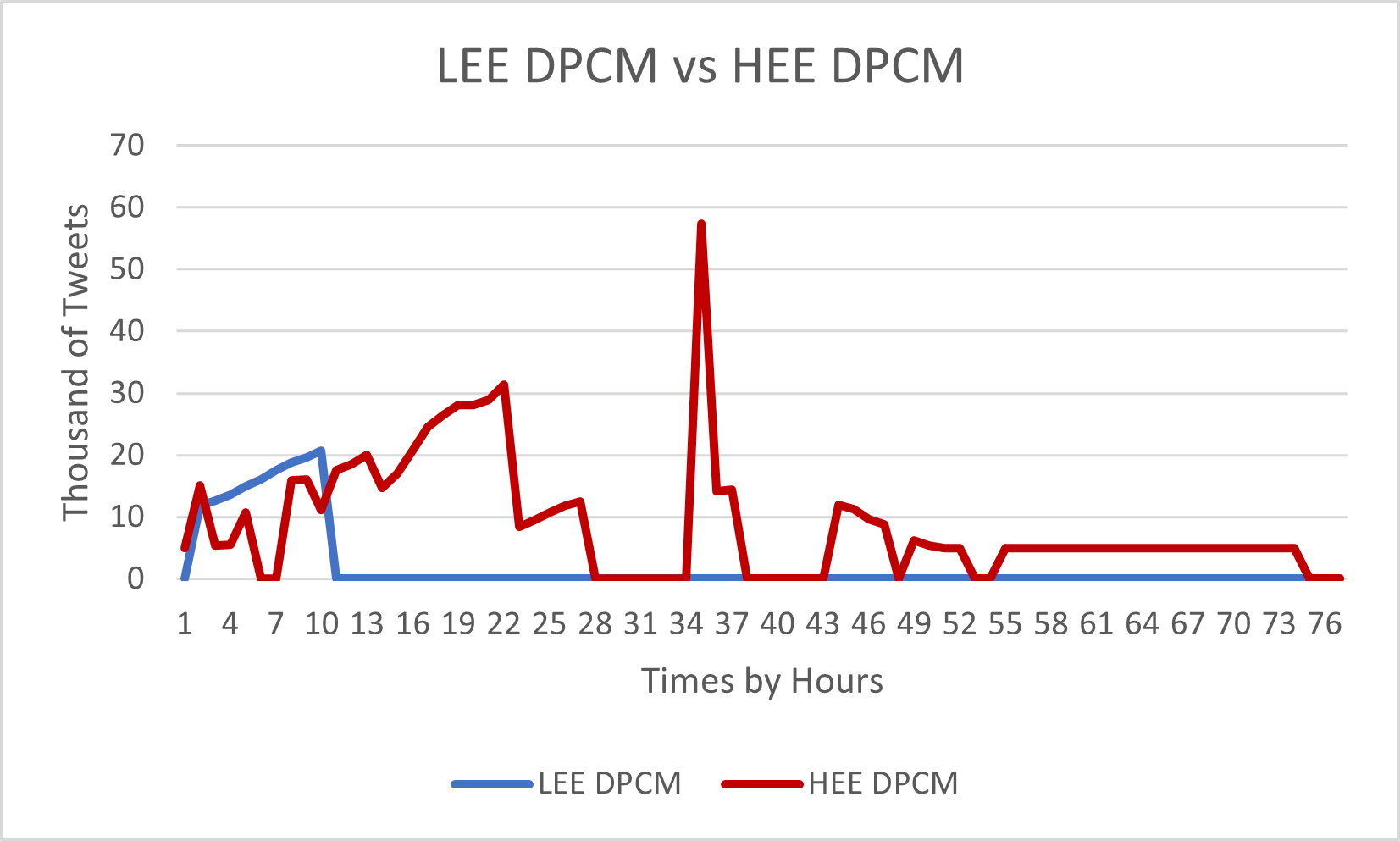}
	\caption{Twitter data trend comparison.}
	\label{FIG:2}
\end{figure}

\subsection{Congiunti \& DPCM} 
Congiunti and DPCM was two important topic and hashtag during the Italian first Covid wave.
The hashtag respects the two phases of the Covid pandemic, the discovery, and the adaptation.

On April 26th 2020, after three months of lockdown, Italians were anxiously waiting for government news about legal procedures to finally meet parents, relatives, friends, and significant ones.  
DPCM is the acronym for "Decree of the President of the Council of Ministers", and it was a legislative way to adapt complex organizations (like government, society, and institutions) to adapt and defend themselves from the Covid pandemic. 
The DPCM's guide from the government wasn't clear enough, the government had used too specific or unknown words for described the ending lockdowns procedures.

The \#Congiunti was indeed the core problem of the information dissemination, because in Italy “Congiunti” (“joint people” in English) does not have a legal definition, so it was interpretable in many ways, like: can an unmarried couple be considered “Congiunti”? What about family members not living together? What about second/third cousins? What about a son/daughter living in separate cities? What about a couple of the same sex (formally cannot be married in Italy). And the list goes on. 
The term shows the highest entropy value, and Italians were largely confused about the word “Congiunti”, and its meaning was essential to properly understand the restrictions. 
Five days after the official announcement, the Italian government published some FAQs to clarify the meaning of the term “Congiunti”. 
As already mentioned, the “Congiunti” case of April 2020 was not the only one we observed and measured.

However during the second wave of Covid, another \#DPCM on October 25th 2020 has brought attention to Italian society, This new DPCM was defining a three-tier system based on a color code according to the intensity of the epidemic danger for the different areas in Italy. This measurement was established to adapt the society and the public administration to reduce the Covid diffusion in different areas, avoiding another hard lockdown procedure. Each of the 20 Italian regions was assigned a different color: Red (high-risk), Orange (medium risk), and Yellow (low-risk) zones. Each color defines a specific set of specific restrictions. 
Italians were confused about which color their region would become, causing limitations to meeting people or social confinement. 

The entropy of this topic became high due not only because of knowing the color of your own region, but also to understand the parameter about how this limitation was implemented. Was more important to determine the colour, the amount of daily Covid cases, or the amount of daily free hospital seats?

\subsection{Beirut}
On 4th August 2020 a large explosion occurred where more than 215 people were killed and about 7,000 people were wounded. All this left the Lebanese capital in ruins. All the world started to wonder how in 2020 that was possible so they started to write on social networks all the possible theories they could think of, this is one of the main reasons why we found an high entropy rate, as people were constantly seeking for new information. 

Many Lebanese people blame security and political officials who failed to enforce safety regulations but until now no one has actually held responsible for that. All we have up to this moment is that the ammonium nitrate was improperly stored at the port of Beirut. All this led to huge protests not only in social networks but also in real life, where what began as a protest outside the Palace of Justice turned into heavy gunfire on the streets.

\subsection{Debates}
The 2020 United States presidential debate was defined by various newspapers like an "unwatchable". The first debate, scheduled on 29 September 2020 was a mess for the chaos it was generated as Trump turned it into chaos as he repeatedly talked over Biden and attacked his Democratic opponent. United States citizens have discussed who was the winner in the midst of all the chaos, and all this situation generated - once again - a mediatic effect and an high entropy rate as irony and anger by people arose in social networks as they were also powered by the classic social media, for example the ABC's journalist George Stephanopoulos define it as the \textit{"worst presidential debate I have ever seen in my life"}. 

The second debate was scheduled for 15 October 2020, but was cancelled due to Trump testing positive to Covid-19. The reaction about this matter on social networks was hilarious as people made fool of Trump saying - someone ironically - that he could not stand the debate so he pretended to be positive. 
The finale debate was scheduled for 22 October 2020 and was - once again - a mess. The debate has been marked by Trump's constant interruptions and insults. For the 60\% of viewers it was the most bottoming out moment of the American political history. 

\subsection{Black lives matters}
On 25th May 2020, George Floyd was murdered by a police officer in Minneapolis, Minnesota. When Floyd was killed, the city became the epicenter of riots. After his death, protests and violences spread out quickly across the United States, against the police. The death of George Floyd gave rise to a series of protests and civil unrest against police brutality and racism. Protests quickly spread nationwide in support of the Black Lives Matter (BLM) movement. 

The movement has been around since 2013, but the death of Floyd arise the protest and the sentiment against police. BLM is a \textit{community organising}, which starts with the support of local communities. Not having a leader means that BLM is at the mercy of disorderly outbursts of social anger: it is organised through a decentralised and horizontal structure. Social media played an important role in the growth of BLM, especially at first. 

We detect an high entropy-value because people wanted to know the motivation for that police violence. Not only for the George Floyd case, but also violence in manifestation. Social media also proved to be much more than an efficient tool to spread the message: they were a way to deepen people's understanding of structural racism by showing connections between seemingly unrelated incidents.


\section{Result \& Newspaper metrics}


Newspaper industry has been changing it is profile during the last decade. As it is still an industry we need to make it clear that they base their main goal is to make money by writing and selling news. For various years this was achieved by being politically polarized and receiving money incomes from both the political sides and the advertisement on the newspapers themselves. While this is still true, we are assisting to a gradual shift of main revenue from the classic paper to the new way news are distributed: online newspapers offer way more information and in a quicker way, for example it was necessary for it to be physically printed daily while this concept is different, there is always a no-stop flow of contents that can be added anytime. Advertisement is now done in a digital way and - usually - as they get a revenue for every viewer that visits their links they tend to write catchy titles that will try to catch as much readers as possible. 
This new way of sharing news is not only useful for final readers but it also allows social researchers to study a phenomenon in an easier way, as thanks to scraping and API techniques, it is possible to extract and analyze them by using digital approaches, reaching the goal to understand social and information dynamics, like HEE and LEE.

\subsection{Information time-patterns during DPCM's events} 
During the 3 months of lockdown, Italian people waited for the government news about the procedures to meet parents, friends, and other people lockdown. We observed that many days before the government’s communications, this topic was already consulted in newspapers, we checked it on "Il Fatto quotidiano" and his site "ilfattoquotidiano.it" (most consulted online newspaper in Italy in 2020).

\begin{figure} 
	\centering
		\includegraphics[width=\linewidth]{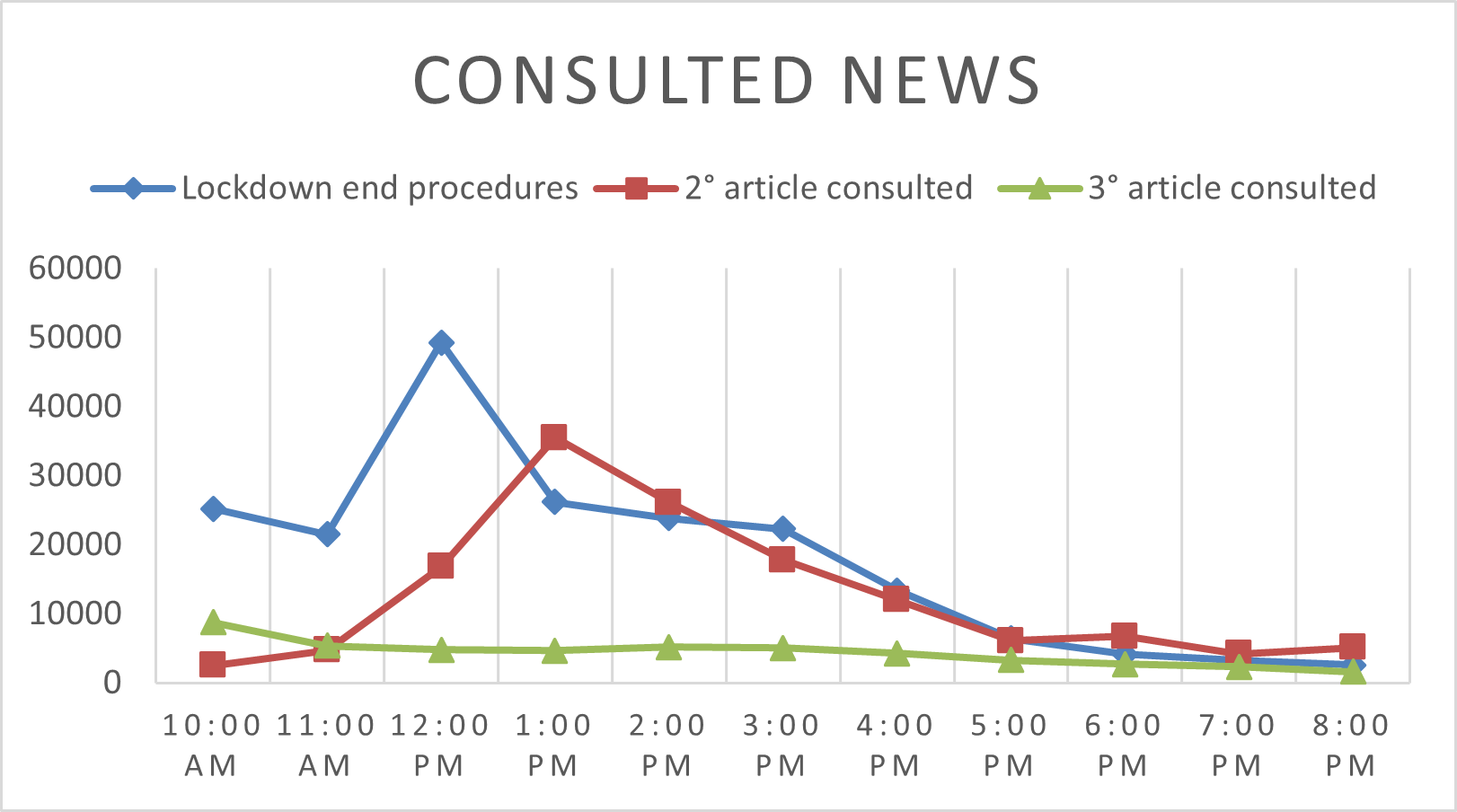}
	\caption{Consulted news Ilfattoquotidiano.it on 26 April 2020}
	\label{FIG:3}
\end{figure}

Data from newspapers shows that there it was a “social hype” about what kind of procedure was followed after the lockdown.
In the days before the ending pandemic procedure, we observed that many people looking for information-topic (Figure \ref{FIG:3}), and the goal of senders (newspapers) was to dispatch the right message to most possible people.
As already prove in the last section, during the institutional communication, some words emitted have a high entropy level and it was difficult to understand properly.

\begin{figure}
	\centering
		\includegraphics[width=\linewidth]{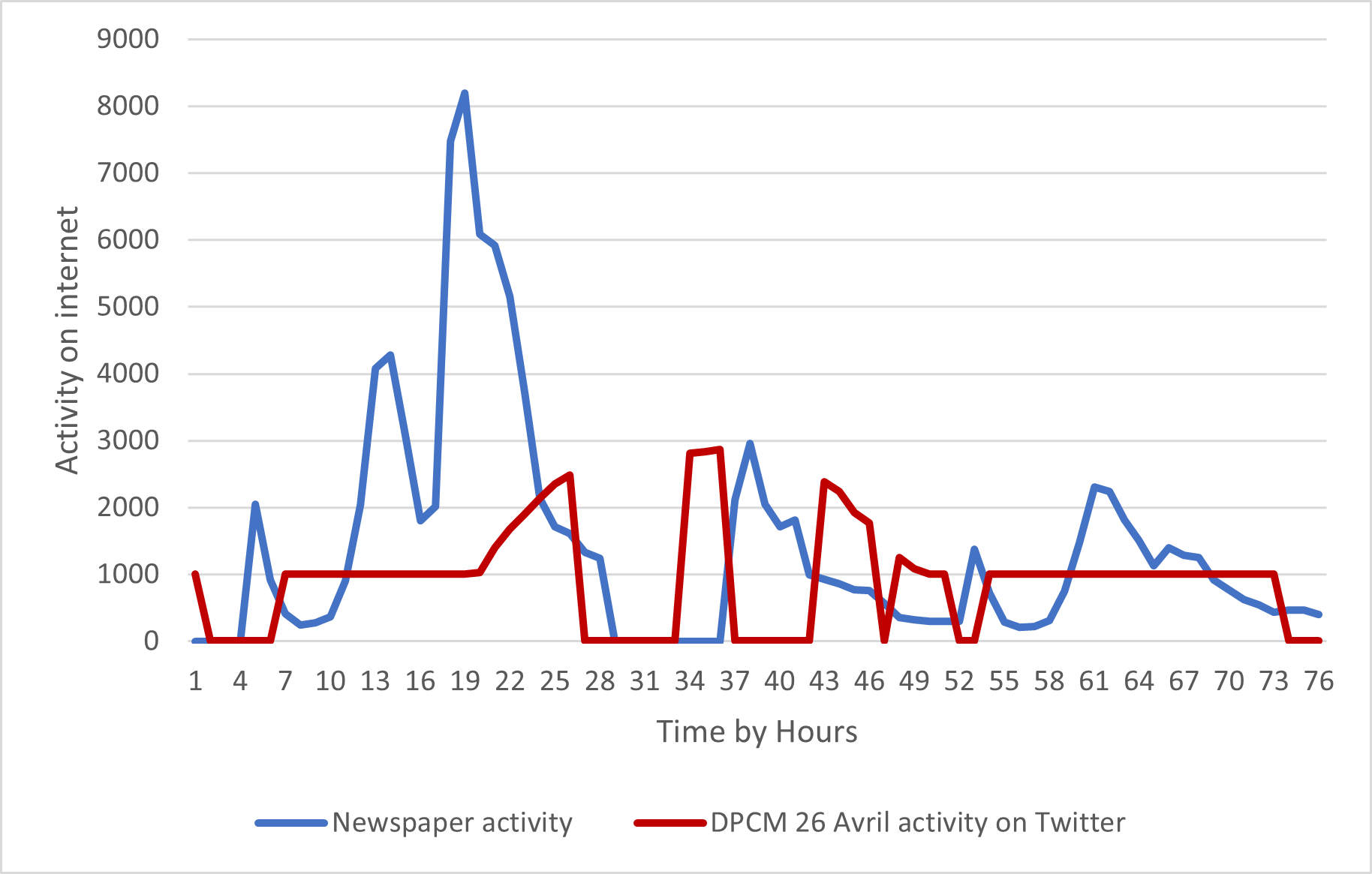}
	\caption{Newspaper activity and SN - 26th April}
	\label{FIG:4}
\end{figure}

In Figure \ref{FIG:4}, we merge this schema with the data gathered from newspapers and social networks, it is possible to infographic what happened on 26 April 2020. The government announcement started at 20 PM, and at this point news (about the interesting topic) in newspapers stop to be seen. It took only 5 days for the government to define what \textit{Congiunti} means.

Figure \ref{FIG:4} shows the activity from social networks and newspapers about the Congiunti topic (DPCM of 26th April).
Early activity on social networks arises before the conference (time 0), evidencing a social hype and huge request from the incoming news related to the incoming information event.
In this case, the social Hype is due to the request for information about the end pandemic procedure.
The Activity on the social network is the Twitter trend data for \#Congiunti Hashtag between 26th April until 1th May 2020, for a total of  $\sim$894.500 tweets.
The Newspaper activity is the visualization of the ending-lockdown-procedures articles from the newspaper-site "ilfattoquotidiano.it", between 26th April until the 1st May 2020, for a total of $\sim$601.930 visualization.
Unfortunately data from ilfattoquotidiano.it doesn't count the refresh of the page from the same unknown account.

From the newspaper's point of view (in figure \ref{FIG:3}), the 38,62\% of the visualisation on the newspaper coming before the Government communication. 
Before this moment, all the visualization rose only to understand the information shared by the government, and the news procedure for the summer holidays. 
And in fact, in Figure \ref{FIG:4}, the curve shows a "social hype" from peoples for the imminent lockdown-ending-procedures.\\

Figure \ref{FIG:4}, show also a curious dynamic from the newspaper's perspective. 
At 10 o'clock we can see a starting big blue curve that emerges after some hour of inactivity (given by night time). This blue curve arise due the night before, the Italian Prime Minister Giuseppe Conte, held a talk (on TV) to the nation, to inform them about the misunderstanding or disinformation about the incoming national procedures after lockdown.
The red line (activity from Twitter) didn't fall during the night, due to the discussion on TV from the Prime Minister, but also because people on the social networks have still continued the discussion about the unclarity of the message from the Prime Minister.
The joint people-case didn't disappear early soon, for many months TV-program and many political talk-show have used the word "congiunti" as a keyword to describe the interaction between people from outside family members, and the possibility to transmit the Covid-19 virus.

\begin{figure} [H]
	\centering
		\includegraphics[width=\linewidth]{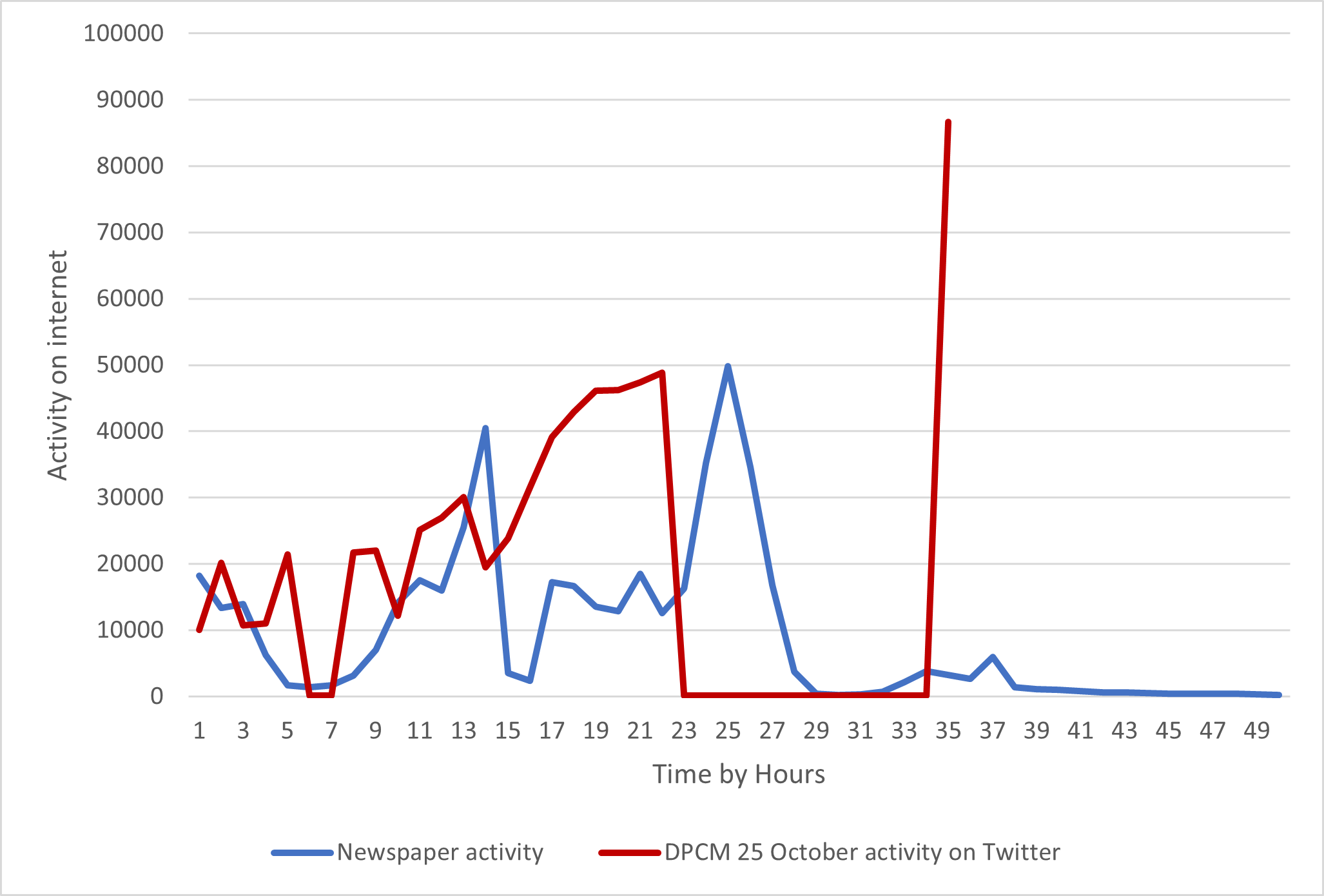}
	\caption{Newspaper activity and SN - 25th October}
	\label{FIG:5}
\end{figure}

In Figure \ref{FIG:5} we show the activity from social network and from newspaper about the DPCM topic (DPCM of 25th October). 
This event don't show any early insight, because on the October 2020 the Italian government publish many DPCMs with very small result to reduce the imminent second wave of Covid pandemic. 
An early insight arise when big change in society are await. But if the expectation for the big change is delayed many times, people stops to bring attention for any news. Some TV program called the publishing of many DPCMs as "born already dead" laws. \\
However, the activity on social network is the Twitter trend data for \#DPCM Hashtag between 24th October and 26th October 2020, for a total of $\sim$618.150 tweet.
The Newspaper activity here is still focused on the early communication process before the other HEE event, showing that people are interest on the topic. They asking information and be informed about the standard parameter to the local regional restriction (as already described on Case study chapter). The visualisation about the article DPCM between 25th October and 28th October 2020 are $\sim$607.725.
As Figure 5 show there is an high density of people reading news in a shot time (8 Houres), as the 20.39\% (123.930) of the total. 
But the must major of readers arise just after the midnight, we count 206.041 visualisation, as the 33.90\% of the total, but distributed over a longer period of time (8 hours for the first, 12 hours for the second - relatively 8H and 28H-44H).

In both case, we saw after 24 hours a total reduction of visualisation on newspaper and the arisen on activity on social network, proving that the information send by the government wasn't clear enough to satisfy the information request from people. 

Thus, we think that the 12 hours of reading by people, is due to the people who wanted to carefully understand what they could do in the various limited zone (Yellow, Orange and Red zone). 
For example understand if it is possible to go out of the region, meet some friend or parents, or go to a restaurant or drink outside.

\begin{figure} [H]
    \centering
    \includegraphics[width=\linewidth]{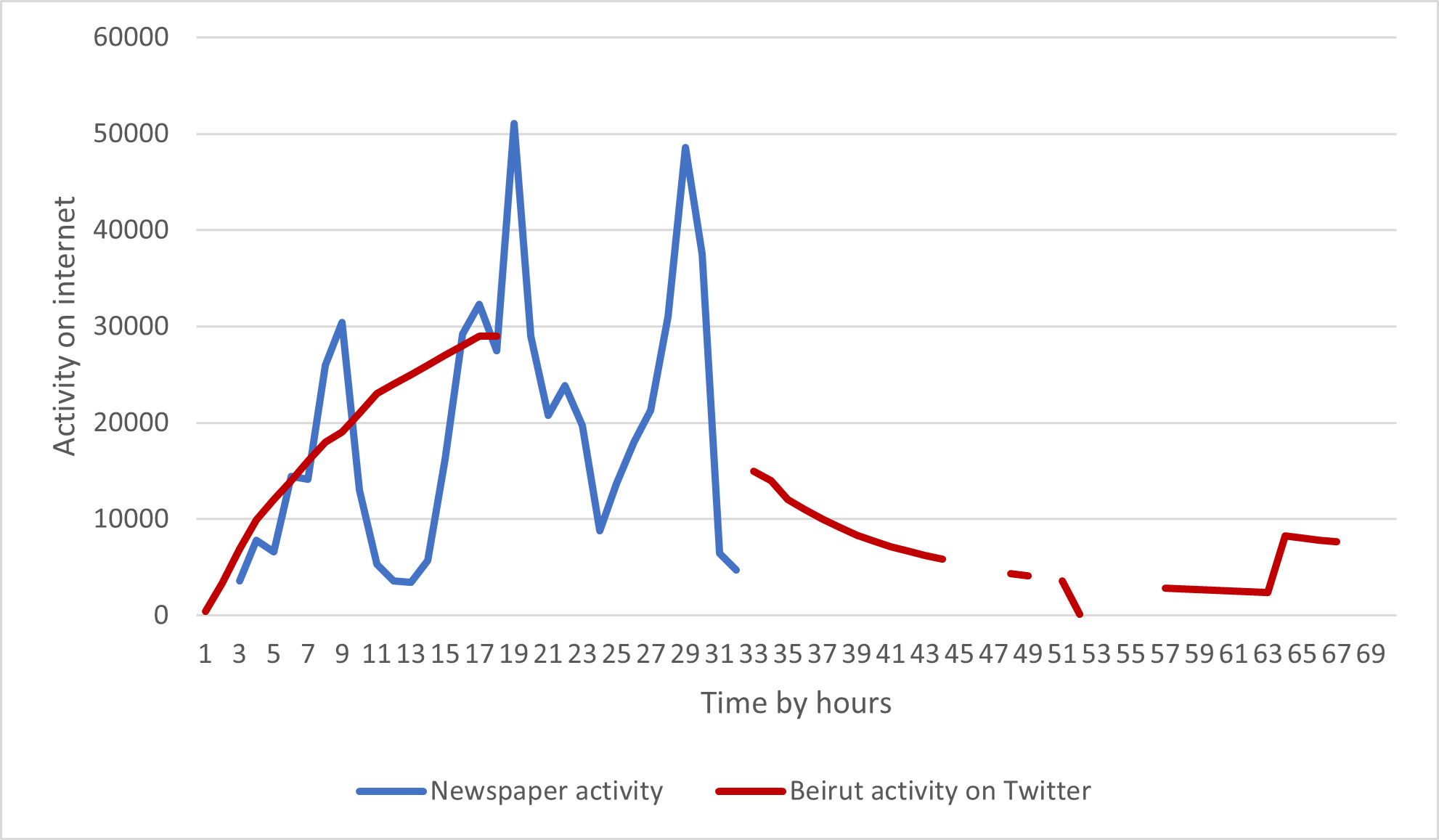}
    \caption{Newspaper activity and SN - Beirut}
    \label{fig:6}
\end{figure}

As shown in the Figure \ref{fig:6}, they are taken into consideration 69 hours (starting from 6 PM o'clock) since the start of the event: during the events occurred in Beirut it can be seen that during the first 7 hours there is almost the same activity in both the newspapers and Twitter, this demonstrates that people were constantly looking for information about the matter. There is a gap of Twitter coverage between the hour 19 and the hour 33, during the same period two huge spikes were registered on newspaper instead. After hour 33 we see no coverage anymore in newspapers, the only events registered after that time is on Twitter, because people were commenting about the previous news on newspapers. As we can see the event started to decline reaching its minimum point in hour 63.
There are some points that can be noticed in the Figure 6, there is a large activity on newspapers that matches the lunch (first spike) and the morning after the event occurred (second spike). 

\begin{figure} [H]
    \centering
    \includegraphics[width=\linewidth]{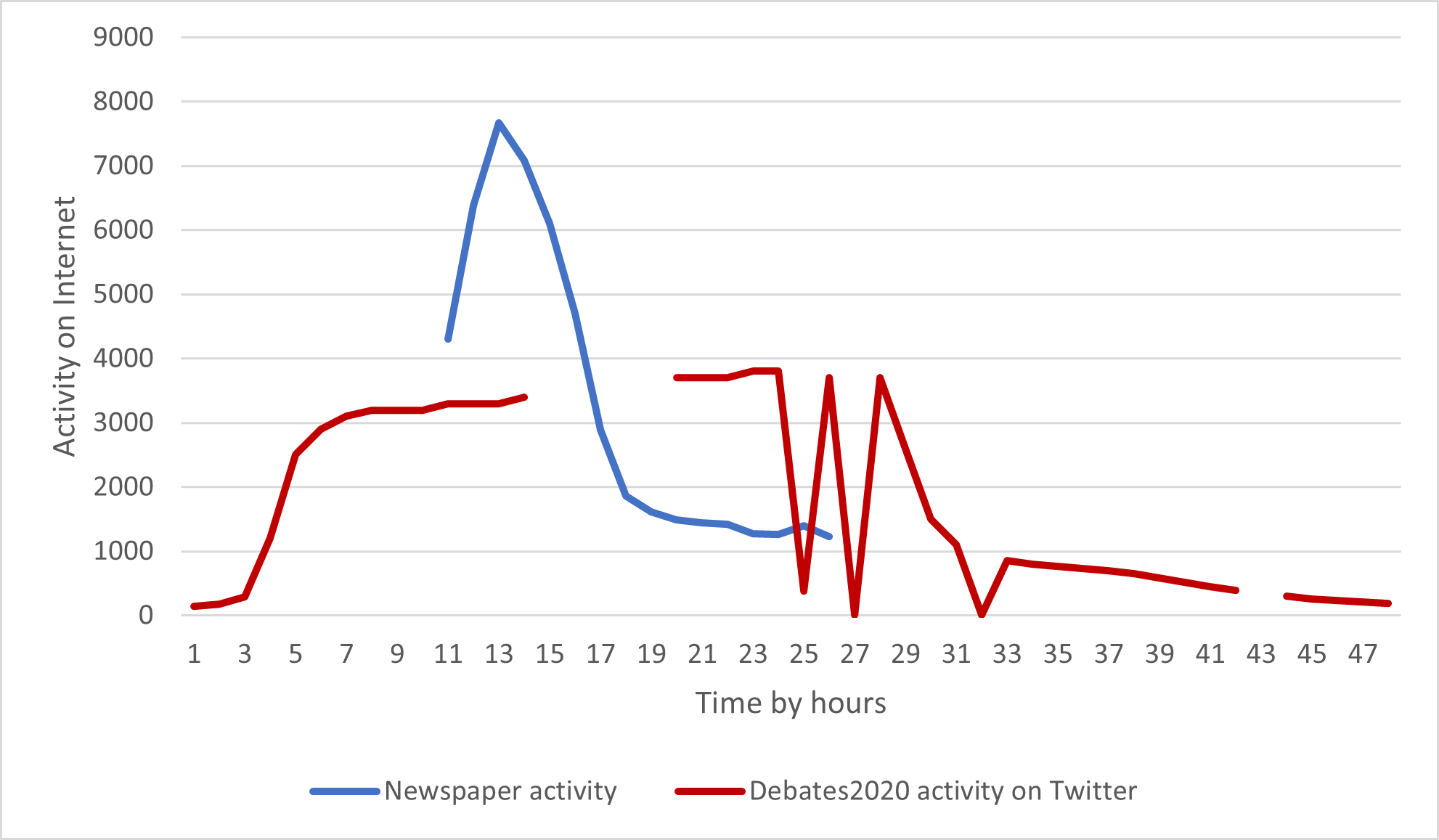}
    \caption{Newspaper activity and SN - Debates2020}
    \label{fig:7} 
\end{figure}

Instead, for the topic Debates in Figure \ref{fig:7}, we can see a greater activity on Twitter than the newspapers, nevertheless they show a linear degree growth that once it reaches its peak it suddenly decreases up to settle to an average value. The activity on Twitter, instead, follows a different pattern, it is way more bound to the topic interest. We can notice some fluctuating peaks until the hour 33, moment since people on Twitter lost interests on the Debate topic.  

\begin{figure} [H]
    \centering
    \includegraphics[width=\linewidth]{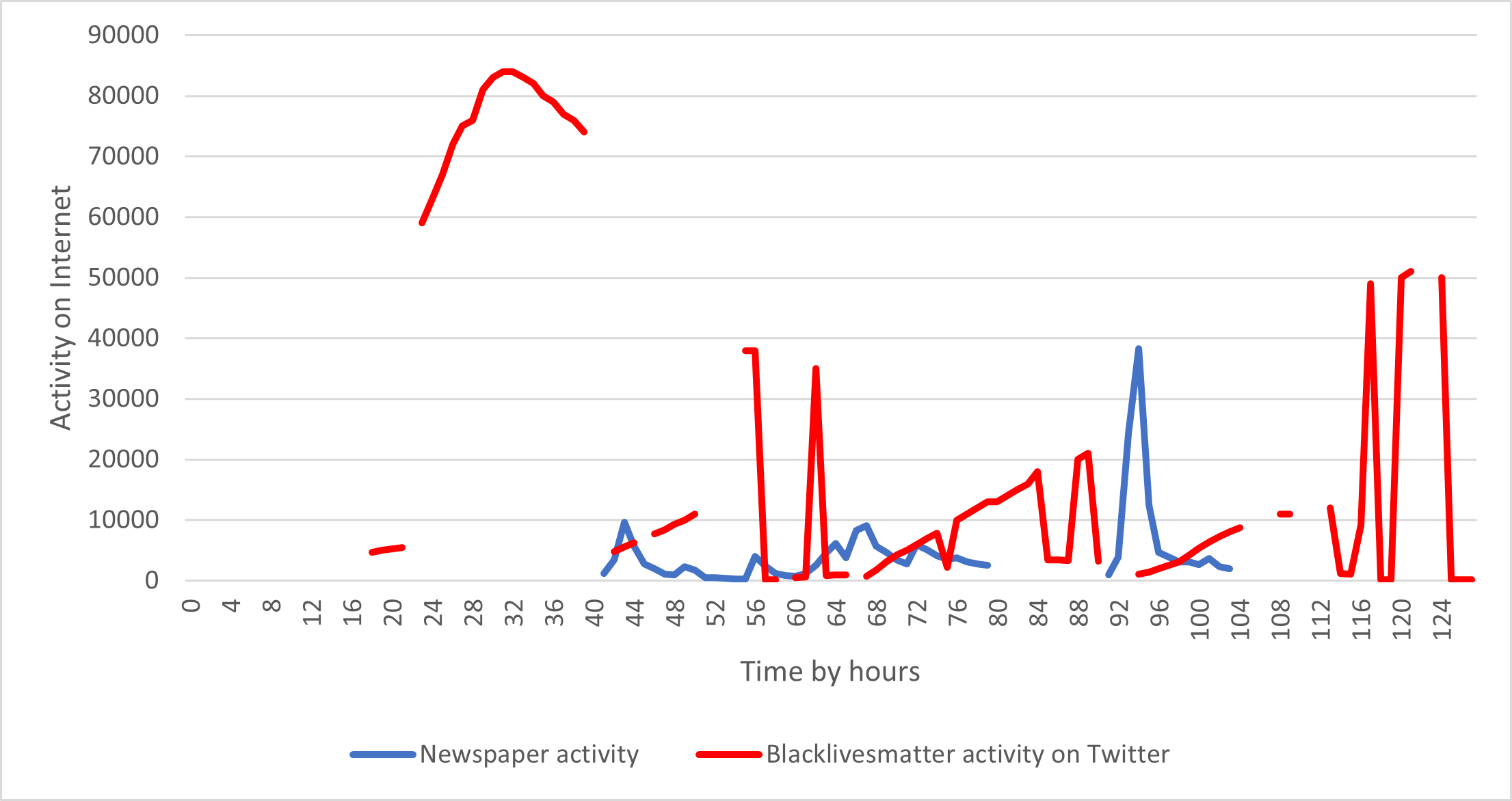}
    \caption{Newspaper activity and SN - BlackLivesMatter}
    \label{fig:8}
\end{figure}

Eventually, the topic BLM is perceived way more by the online community than the newspapers. In the first hours of the event show an intense activity on Twitter reaching a maximum peak of about 83k; newspapers - instead - tend to delay the beginning of the focus, it was registered on them only after 40 hours that the event was already being discussed on Twitter. Systematically, it is preferred to search for this kind of info on Twitter than the newspapers, as they never reach the same level of activity of Twitter but only about the hour 94 after the start of the event as shown in the Figure \ref{fig:8}.

We can clearly notice that after a red peak there is an increasing number of readers that is translated in an increased number of views online. These peaks can also be seen more than once during the same day for the same event, highlighting a possible correlation between 1) peaks of tweets; 2) need for additional info (people move to newspapers to get informed); 3) growth of number of views of the online newspapers.


\subsection{Dynamic prediction about newspapers}

HEE are very rare events, but otherwise, they show the unbalance between the information requested and information supplied by the senders.\\ How is it possible to estimate if something is wrong during the information communication? and how is possible to expect High Entropic Events and predict online visualization on newspaper?  
We have enumerate the various action steps in information dynamics, trying to highlight the crucial moments and which moments may be most significant for prediction the online activity for newspaper and social network. 

\begin{enumerate}
    \item If the information request is really important or awaited, it usually shows some insight before the moment that provides that information needed
    \item The moment that everyone listens and tries to assimilate the information, is a crucial stage because if the communication goes well (good information diffusion and right information received) there is no HEE (only LEE), but if the entropy of the message/information is high, the HEE is arisen
    \item After that phase, people usually start to debate and talk on social networks and/or in their private circles (family, friends, etc..). Thus, at this point, if enough people start to talk, the topic grows and becomes a Trend topic for a long time (at least 40 hours). At this point, we can evidence how much the information diffusion was spread and how much the information was clear 
    \item The newspaper’s goal is to make money by selling news. The newspaper strategically must talk about that topic and they start to write some articles to supply the right information, but if the information’s entropy was too huge, this also affects the newspapers, and they cannot supply the right information (because they also didn't understand the information because was to entropic or because the topic is too complex, and it required time to be assimilated). People usually continue to talk for at least 2 days
    \item If was an organization that shared the information, at some point, the organization that has to provoke the HEE, should dispatch the right definition of the information sends 
\end{enumerate}

The rise and climb in any HEE figure, enlighten the hypothesis of some time-pattern during the case-study event, and by reaction, a possibility to quantify and predict online visualization between social network and newspaper. 
After the big peak activity on social network, a peak activity also pops up in newspapers after a short time. This dynamics is enlighten in every HEE, but we didn't found the same dynamic from LEE.
To quantify the correlation activity from social network and newspaper we have calculate in Table \ref{tab:3}, how much time in average the first article as been create after the peak on social network, and how much time take the newspaper to obtain a huge amount of visualization to the article. 

However, for some topic like BLM or Beirut the average is higher then the others topic, because of the time zone where the events occurred. That means that if the event arise in the United state, people in EU are still sleeping, by consequence it hard to writhe and diffuse online news. 
Indeed, people usually read newspaper in early morning or during the afternoon for inform about some specific news;  moreover, about LEE, the activity on social media arise in early morning because is related to specific news of the day (\#giornatamondialedellibro,  \#mothersday), or also when some specific event arise (\#GazaUnderAttack, \#Superlague, footballs match and others).
Another important piece of information that newspapers can obtain, from the HEE prediction forecast, is the number of newspaper articles. Still in Table \ref{tab:3}, we show the higher net value of number of HEEs articles than LEEs. 
HEE events find more articles in newspapers (to explain the event and/or its evolution), while LEEs have circa only one article.
\begin{table}[ht]
    \small\sf\centering 
    \caption{Difference time between: Twitter-article and mayor increase by hours.}
     \label{tab:3}
    \begin{tabular}{l|l|l|l|l|c}
    \toprule
 Beirut &	BLM	& DPCM (Apr) & DPCM (Oct)& Debate &	TOT\\
 \midrule
3.33 &	7.75&	3.75&	1.25& 3&	4,77 \\
1.67&	3&	2.75&	1&2&	2.60\\
\bottomrule
    \end{tabular} \\[10pt]
    \begin{tabular}{lc}
\toprule
Entropic event&Average articles\\
\midrule
\texttt{HEE}&\textbf{ $\sim$ 6.4} \\
\texttt{LEE}&\textbf{$\sim$ 1}\\
\bottomrule
\end{tabular}
\end{table}

\subsection{Information structure}
During the data analysis we also observed a different structure of the messages \cite{R26} relative to HEE and LEE.
We observed that HEE have on average more question mark then LEE, consequentially to proof this result we research and gathering data from others HEE and LEE (\#MilanNapoli, \#Giornatamondialedellibro, \#Gazaunderattack etc..). As Table \ref{tab:4} illustrate, HEE response to the events are structurally different from LEE. The average indicates the probability of any tweet to contain a question mark. 
The data suggest a different information structure about the social reactions to some topic, proving that HEE and LEE are quite different not only in the "natural decay" of the topic, but also in the reaction structure.

\begin{table}[ht]
\small\sf\centering
\caption{Question marks on HEE and LEE.\label{T1}}
\begin{tabular}{lllll}
\toprule
Topics&Tweets&Q. marks&Events&Average\\
\midrule
 Debates2020 & 868 & 210 &HEE& \textbf{24.2\%}\\
 DPCM (Apr 26th) & 2000 & 312 &HEE& \textbf{15.6\%}\\
Megxit & 115* &  17 & HEE& \textbf{14.8\%}\\
 BLM & 1612 & 154 &HEE& \textbf{9.5\%}\\
 DPCM (Oct 25th) &2000& 178 &HEE& \textbf{8.9\%}\\
  Beirut & 724 & 53 &HEE& \textbf{7.3\%} \\
 & & & & \\
 Valentinesday &1600& 251 &LEE& \textbf{14.25\%} \\
 JuveInter &1000& 71&LEE& \textbf{7.1\%}\\
 G.dellibro &1000& 69&LEE& \textbf{6.9\%}\\
SuperLega &1000& 68&LEE& \textbf{6.8\%}\\
Int.WomensDay &1000& 64&LEE& \textbf{6.4\%}\\
 MilanNapoli &1000& 61&LEE& \textbf{6.1\%}\\
 DPCM &1000& 60&LEE& \textbf{6\%}\\
 SuperLeague &1000& 55&LEE& \textbf{5.5\%}\\
 MothersDay &1000& 31&LEE& \textbf{3.1\%}\\
 GazaU.Attack &1000& 14&LEE& \textbf{1.4\%}\\
 Dupasquier &1000& 14&LEE& \textbf{1.4\%}\\
\bottomrule
\end{tabular}\\[10pt]
\begin{tabular}{lc}
\toprule
Entropic event&Average\\
\midrule
\texttt{HEE}&\textbf{13.38\%} \\
\texttt{LEE}&\textbf{5.90\%}\\
\bottomrule
\end{tabular}
 \label{tab:4}
\end{table}

As shown in Table \ref{tab:5}, five semantic areas were identified. Keywords were identified for each area.
We have use Odds-ratio to prove if there is a topic connection between the topic from social network, and online news from newspapers. 
The methodology of the odds-ratio, needs to clean and set the data in a good condition for the analysis.
\begin{table}[ht]
    \small\sf\centering 
    \caption{Compatibility between Tweets and Words.}
    \begin{tabular}{l|c|c}
    \toprule
 Topics  &	N. of articles & Odds \\
 \midrule
Congiunti& 8 &	0,132\\
Beirut& 3 &	0,081\\
Debates& 2 &	0,063\\
BLM& 3 &	0,047\\
DPCM& 3 &	0,031 \\
\bottomrule
    \end{tabular} \\[10pt]
    \begin{tabular}{cc}
\toprule
Average Articles & Average Odds\\
\midrule
\textbf{ $\sim$ 3.8} & \textbf{ 0,071}
\end{tabular}
 \label{tab:5}
\end{table}

For example, due to grammatical reasons, documents are going to use different forms of a word, such as "organize", "organizes", and "organizing". Additionally, there are families of related words with similar meanings, such as democracy, democratic, and democratization that share the same prefix.

To reduce the dimensionality of the words, we performs a lemmatization analysis, that is a text normalization techniques in the fiels of Natural language Processing that are used to prepare text, words, and documents for further processing (i.e. the lemma of "are" is "be", the lemma of "debates" is "debate" and so on).
Afterward, we cross-checked words in articles and tweets by counting the occurrences for each one of them in both of the sources. This allowed us to calculate an article/tweet compatibility index for each topic. 

Finally, we calculate the odds ratio in order to evaluate the correlation between articles and tweets. We found that, for the equal N. of articles, some topics show more interest in online communities pushing users to discuss it on online platform.


\section{Conclusions}

In a world where the communication is becoming faster and faster, also thanks to the use of the ICT technologies, social media are widely used for communicating information of various kinds \cite{R14}. 

Blogs, social networks, and online newspapers are now social relationship tools and they build weak and strong ties. In the traditional world, newspapers simply had to give out information, and people would consume it by reading or looking at it without interacting. All social media are not only used to communicate but they are a new way to stay informed and share thoughts about anything, like if you were in a public place talking to someone.  

Simply making information available is not enough for today’s public. Today’s audiences expect to be able to choose what they read, and most believe they should be able to contribute content and opinions, too. In particular, the one-to-many nature of Twitter created an opening for governments to disseminate relevant messages \cite{R16}. 

When news breaks – as in our case study – people increasingly turn to the Internet. Social media platforms became powerful tools for communicating rapidly and without intermediary gatekeepers. As we have observed, the quality of information communicated through digital platforms is essential to prevent misunderstandings and confusion about the message. 

The more controversial a topic or message is, the more significant people's reaction on social media and the higher the possibility that, as a result, misinformation will be created on the subject. We introduce an entropic model allowing researchers to:

\begin{itemize}
    \item catch how much people have understood about a relative news on social network
    \item discover a different information structure about different reaction by people, and different "natural decay" of the information
    \item proving  better strategy and tactic about information diffusion
    \item showing high entropic social event in  society
    \item to transmit an information at an optimal level it is suggested for the entropy to be slightly above the average between high and low entropy, this way news will be easily understandable by the vast majority of people and at the same time, alluring to the peoples.
    \item show methodology to predict interaction/visualization from social network to newspaper.
\end{itemize}

Social media can perpetuate misinformation about many themes, for example, on scientific questions such as antimicrobial resistance, and may contribute indirectly to the misuse of antibiotics \cite{R17}. Thus, it is critical that when a message is delivered, there should not be miscommunication between the sender and the receiver. To avoid misunderstanding, it must be sure that the message is clear and straightforward, and counting Questions marks on the comments about that topic could be a valuable method to rapidly identify doubts. 

The clearer and more understandable the message, the lower the observed entropy level (LEE). The lower the ability to spread clear and comprehensible messages, the higher the entropy (HEE) level, thus expanding social media's potential role in conveying misinformation and hampering the online newspaper's work to deliver news and information. 

Social media nowadays represent valuable, informative aid for the research. By adding an entropy analysis to the classification of social events, it was possible to analyze the miscommunication process by senders, showing the level of diffusion of information. We observed that if people talk a lot about some specific topic for many days, they unconsciously need more explication and information. The case study showed how high entropy events (HEE) and low entropy events (LEE) are related to specific behavior on social media. Usually, the "natural decay" of LEE (predictive and routines events like football matches) is lower than five hours, showing standard behavior and a balance right information assimilation. 

Conversely, HEE are in general unpredictable: they trigger a need for additional information, leading to complex behavior that can be observed on social media. HEE seems to provides an exciting research challenge.

\subsection{Author Contributions}
Investigation, Data resources A.R.; Data elaboration, A.R., A.P. and V.M.; Methodology, F.M.; Data cleaning and Software, A.R. and G.G.; Correction F.M. and G.G.\\
All authors have read and agreed to the published version of the manuscript.

\subsection{Copyright}
Copyright \copyright\ \volumeyear\ SAGE Publications Ltd,
1 Oliver's Yard, 55 City Road, London, EC1Y~1SP, UK. All
rights reserved.

\subsection{Funding}
The author(s) disclosed receipt of the following financial support for the research, authorship, and/ or publication of this article:
This project has received funding from the University of Catania. 

\subsection{Author biographies}
\orcid{0000-0003-3816-0539} Andrea Russo is a Ph.D. candidate in Complex Systems at the University of Catania. 
He is currently working at the Department of Physics and Astronomy. 
His main research field and interests are focused on the study and the development of Computational social method to explain and research social complexity, particular field like Politics, Economics, Business and Defense-Security sector applications. \\

\orcid{0000-0001-8777-6626} Antonio Picone is a Ph.D. candidate in Complex Systems at the University of Catania. His main project and his interests are bound to the field of Natural Language Processing, with a particular interest in researching ways to identify sentiments and emotions in a written text with the help of Artificial Intelligence.\\

\orcid{0000-0001-7834-5189} Vincenzo Miracula is a PhD candidate in Complex Systems at the University of Catania. He currently works at the Department of Physics and Astronomy. His research interests are in computational social sciences, artificial intelligence and natural language processing, with a particular interest in network theory, text analysis and spreading fake news.\\

\orcid{0000-0003-3847-2372} Francesco Mazzeo Rinaldi is an Associate Professor at the University of Catania, where he teaches evaluation research methodology. He is also an Affiliate Professor at the KTH, Royal Institute of Technology, School of Architecture and the Built Environment, Stockholm. In the last 15 years, he carried out training, research, and consulting activities on program and policy evaluation, monitoring systems, and big data analytics in several regional, national and international public organizations in social, development, and cohesion policy.\\

\orcid{0000-0001-5490-779X} Giovanni Giuffrida is researcher at the university of Catania.
He obtains his Ph.D at UCLA (Los Angeles) in computer science.
CEO of Neodata group SRL, Founder and CTO of Data Corporation, Santa Monica, California.\\



\bibliographystyle{SageV}

\bibliography{biblio.bib}

\begin{thebibliography}{10}
\providecommand{\url}[1]{\texttt{#1}}
\providecommand{\urlprefix}{URL }
\expandafter\ifx\csname urlstyle\endcsname\relax
  \providecommand{\doi}[1]{DOI:\discretionary{}{}{}#1}\else
  \providecommand{\doi}{DOI:\discretionary{}{}{}\begingroup
  \urlstyle{rm}\Url}\fi
\providecommand{\eprint}[2][]{\url{#2}}

\bibitem{R1}
Moberg.
\newblock Mediatization and the technologization of discourse: Exploring
  official discourse on the-internet and information and communications
  technology within the evangelical lu-theran church of finland.
\newblock \emph{New Media \& Society} 2018; : 515–531.

\bibitem{R2}
Guo L and McCombs M.
\newblock Toward the third level of agenda setting theory: A network agenda
  setting model.
\newblock \emph{Annual convention of the Association for Education in
  Journalism} 2011; .

\bibitem{R3}
Bode L.
\newblock Political news in the news feed: Learning politics from social media.
\newblock \emph{Mass Communication and Society} 2016; 19: 24–48.

\bibitem{R4}
Russell~Neuman W, Guggenheim L, Mo~Jang S et~al.
\newblock The dynamics of public attention: Agenda-setting theory meets big
  data.
\newblock \emph{Journal of Communication} 2014; 64(2): 193--214.

\bibitem{R5}
Feezell JT.
\newblock Agenda setting through social media: The importance of incidental
  news exposure and social filtering in the digital era.
\newblock \emph{Political Research Quarterly} 2018; 71(2): 482--494.

\bibitem{R6}
Kaplan HM A.
\newblock Users of the world, unite! the challenges and opportunities of social
  media.
\newblock \emph{Business Horizons} 2010; 53: 59–68.

\bibitem{R7}
Castronovo~C HL.
\newblock Social media in an alternative marketing communication model.
\newblock \emph{Journal of Marketing Development and Competitiveness} 2012; 6:
  117–131.

\bibitem{R8}
Huang L.
\newblock Social contagion effects in experiential information exchange on
  bulletin board systems.
\newblock \emph{Journal of Marketing Management} 2010; 26: 197–212.

\bibitem{R9}
Newell DA R.
\newblock Meeting the climate change challenge (mc3): the role of the internet
  in climate change re-search dissemination and knowledge mobilization.
\newblock \emph{Environmental Communication} 2015; 9: 208–227.

\bibitem{R10}
Murphy SS G.
\newblock Using social media to facilitate knowledge transfer in complex
  engineering environments: a primer for educators.
\newblock \emph{European Journal of Engineering Education} 2013; 38: 70–84.

\bibitem{R34}
Giuffrida G, Mazzeo~Rinaldi F and Russo A.
\newblock Analyzing communication broadcasting in the digital space.
\newblock In \emph{International Conference on Machine Learning, Optimization,
  and Data Science}. Springer, pp. 518--530.

\bibitem{R11}
Mazzeo~Rinaldi F, Russo A and Giuffrida G.
\newblock Information balance between newspapers and social networks.
\newblock \emph{CARMA 2020: 3rd International Conference on Advanced Research
  Methods and Analytics} 2020; .

\bibitem{R30}
Kleeman R, Majda AJ and Timofeyev I.
\newblock Quantifying predictability in a model with statistical features of
  the atmosphere.
\newblock \emph{Proceedings of the National Academy of Sciences} 2002; 99(24):
  15291--15296.

\bibitem{R31}
Li R, Zhao Z, Zhou X et~al.
\newblock The prediction analysis of cellular radio access network traffic:
  From entropy theory to networking practice.
\newblock \emph{IEEE Communications Magazine} 2014; 52(6): 234--240.

\bibitem{R32}
Song C, Qu Z, Blumm N et~al.
\newblock Limits of predictability in human mobility.
\newblock \emph{Science} 2010; 327(5968): 1018--1021.

\bibitem{R33}
Ding G, Wang J, Wu Q et~al.
\newblock On the limits of predictability in real-world radio spectrum state
  dynamics: From entropy theory to 5g spectrum sharing.
\newblock \emph{IEEE Communications Magazine} 2015; 53(7): 178--183.

\bibitem{R21}
Shan S and Lin X.
\newblock Research on emergency dissemination models for social media based on
  information entropy.
\newblock \emph{Enterprise Information Systems} 2018; 12(7): 888--909.
\newblock \doi{10.1080/17517575.2017.1293300}.
\newblock \urlprefix\url{https://doi.org/10.1080/17517575.2017.1293300}.
\newblock \eprint{https://doi.org/10.1080/17517575.2017.1293300}.

\bibitem{R22}
Yin J, Karimi S, Lampert A et~al.
\newblock Using social media to enhance emergency situation awareness.
\newblock In \emph{Twenty-fourth international joint conference on artificial
  intelligence}. p.~1.

\bibitem{R23}
Peng S, Yang A, Cao L et~al.
\newblock Social influence modeling using information theory in mobile social
  networks.
\newblock \emph{Information Sciences} 2017; 379: 146--159.

\bibitem{R24}
Kolli N, Balakrishnan N and Ramakrishnan K.
\newblock On quantifying predictability in online social media cascades using
  entropy.
\newblock In \emph{Proceedings of the 2017 IEEE/ACM International Conference on
  Advances in Social Networks Analysis and Mining 2017}. pp. 109--114.

\bibitem{R25}
Borge-Holthoefer J, Perra N, Gon{\c{c}}alves B et~al.
\newblock The dynamics of information-driven coordination phenomena: A transfer
  entropy analysis.
\newblock \emph{Science advances} 2016; 2(4): e1501158.

\bibitem{R27}
Barros PH, Cardoso-Pereira I, Loureiro AA et~al.
\newblock Event detection in social media through phase transition of bigrams
  entropy.
\newblock In \emph{2018 IEEE Symposium on Computers and Communications (ISCC)}.
  IEEE, pp. 1--6.

\bibitem{R28}
Park HW.
\newblock Mapping election campaigns through negative entropy: Triple and
  quadruple helix approach to south korea’s 2012 presidential election.
\newblock \emph{Scientometrics} 2014; 99(1): 187--197.

\bibitem{R29}
Senevirathna C, Gunaratne C, Rand W et~al.
\newblock Influence cascades: Entropy-based characterization of behavioral
  influence patterns in social media.
\newblock \emph{Entropy} 2021; 23(2): 160.

\bibitem{R12}
Labs NB.
\newblock A mathematical theory of communication.
\newblock \emph{The Bell System Technical Journal} 1948; 27.

\bibitem{R13}
SHANNON CE.
\newblock A mathematical theory of communication.
\newblock \emph{Board of Trustees of the University of Illinois} 1049; .

\bibitem{R18}
Bernikova O, Granichin O, Lemberg D et~al.
\newblock Entropy-based approach for the detection of changes in arabic
  newspapers’ content.
\newblock \emph{Entropy} 2020; 22(4): 441.

\bibitem{R20}
Tang M and Mao X.
\newblock Information entropy-based metrics for measuring emergences in
  artificial societies.
\newblock \emph{Entropy} 2014; 16(8): 4583--4602.
\newblock \doi{10.3390/e16084583}.
\newblock \urlprefix\url{https://www.mdpi.com/1099-4300/16/8/4583}.

\bibitem{R19}
Kim M, Newth D and Christen P.
\newblock Modeling dynamics of diffusion across heterogeneous social networks:
  News diffusion in social media.
\newblock \emph{Entropy} 2013; 15(10): 4215--4242.

\bibitem{R26}
Mazzeo V, Rapisarda A and Giuffrida G.
\newblock Detection of fake news on covid-19 on web search engines.
\newblock \emph{arXiv preprint arXiv:210311804} 2021; .

\bibitem{R14}
Huang L, Clarke A, Heldsinger N et~al.
\newblock The communication role of social media in social marketing: a study
  of the community sustainability knowledge dissemination on linkedin and
  twitter.
\newblock \emph{Journal of Marketing Analytics} 2019; 7(2): 64--75.

\bibitem{R16}
Olteanu A, Vieweg S and Castillo C.
\newblock What to expect when the unexpected happens: Social media
  communications across crises.
\newblock \emph{ACM conference on computer supported} 2015; : 994--1009.

\bibitem{R17}
Andersen B, Hair L, Groshek J et~al.
\newblock Understanding and diagnosing antimicrobial resistance on social
  media: a yearlong overview of data and analytics.
\newblock \emph{Health communication} 2019; 34(2): 248--258.

\end{thebibliography}

\end{document}